# Observations of Manifestations of Skeletal Structures of a Filamentary Matter on the Sun

V. A. Rantsev-Kartinov

INF RRC "Kurchatov Institute", Kurchatov Sq. 1, Moscow, 123182, Russia,
Tel.: 7 (495) 196 7334, Fax: 7 (495) 943 0073, E-mail: rank@nfi.kiae.ru

**Abstract.** In the given paper the author makes some declarative applications:

i) both on the Sun and in its nearest environment they are existing a skeletal structures which topology is identical to those which have already been revealed by the author earlier in a wide range of scales (covering region from $10^{-7}$ cm up to $10^{28}$ cm, i.e., almost of 35 orders of value), phenomena and environments;

ii) the majority of phenomena concerned with powerful coronal mass ejections of the Sun and powerful protuberances are provoked by penetration into an atmosphere of the Sun and its body of an almost invisible coaxial-tubular filaments (may be, of some filamentary matter), and not in result of powerful internal nuclear explosions;

iii) Sun spots very often have three-dimensional constructions, therefore the author offer a hypothesis - Sun spots are a result of pressing out of structures of filamentary matter of the Sun (during its activity) from within to its surface which have the same coaxially-tubular skeletal topology, as the skeletal structures mentioned above;

iv) structures of a filamentary matter (revealed in result of using of the method of multilevel dynamical contrasting for analysis of the Sun images) with radius of their rotation around of solar axis essentially smaller than radius of solar disk on breadth of their observation can testify about formation and carrying-out their images from within star into the external side by means of interaction of the electronic flux neutrino with this hypothetical filamentary matter of the Sun and development of this image (coded into neutrino flux) as image in soft x-ray as a result of interaction of this neutrino flux with rare structure of the same filamentary matter located between the Earth and the Sun.

## 1. Introduction

The author's studies by skeletal structures have started from analysing the photographic images of plasma with the help of the method of multilevel dynamical contrasting (**MMDC**), which was developed earlier [1a, 1b] and is based on the variable computer-made contrasting of an image. An analysis of databases of photographic images of the Sun surface, its atmosphere and the nearest cosmic surrounding, obtained at various magnifications, and for various types of the Sun surface radiation by means of this method revealed the presence of skeletal structures of the Sun (**SSS**) as on the Sun so in its surrounding [2]. The **SSS** topology appears to be identical to that which have been formerly found in a wide range of length scales (as much as about 35 orders of values), environments and for various phenomena in laboratory, Earth atmosphere and space [1, 3]. The typical **SSS** consists of separate identical blocks which are linked together and forms a network. Two types of such blocks are found: (i) coaxial tubular structures (**CTS**s) with internal radial bonds, and (ii) cartwheel-like a structures (**CWS**s), located either on an axle or in the butt-ends of **CTS** block. The revealed **SSS** have a whole series of remarkable properties which have been described before [1c-1f] (See Fig.1and Fig. 2). Such, a long filaments consist of straight ("rigid") nearly identical the **CTS** blocks which are joined flexibly similarly to joints in a skeleton. It is assumed such joints may be realized due to stringing of the individual **CTS** blocks on common the magnetic field flux which penetrates the whole of such filament, and itself the **CTS** blocks are an interacting magnetic dipoles with micro-dust skeletons, which are immersed into plasma.

**Fig. 1.**

**Fig. 2.**

## 2. Coaxial-Tubular Structures and Cartwheel-Like Structures of the Sun

The **SSS** are revealed in the images obtained in different range of lengths of waves. The very large **CTS**s are sometimes observed (see Fig. 3, 4). These structures do not rotate, but they create such configuration of electromagnetic fields in which plasma can have rotation. Therefore the rotary movement of plasma in such case appears as a result of electro-dynamic processes caused by the skeletal structures which are immersed into plasma, but not as a result of development of instability in plasma.

**Fig. 3.**

**Fig. 4.**

## 3. Skeletal Structures in Coronal Mass Ejections of the Sun

The analysis of dynamics of images of large-scale coronal mass ejections (**CME**) of the Sun has led the author to a hypothesis, that explosions on the Sun, basically, are provoked by penetration from the nearest space environment into its atmosphere and body of large-scale almost transparent filaments. Here the hypothesis is put forward, that these filaments belong to structures of a filamentary matter (**FM**) located in the nearest environment of the Sun. In the given paper the problems of nature definition of the observable **FM** and mechanisms of their interactions with the Sun and its atmosphere are not consider, and only the revealed confirming examples of this hypothesis (in the author opinion) are given. The star reacts to intrusions of individual filaments of such the **FM** by powerful ejection of mass of coronal plasma. This plasma at flowing around that filament illuminate from beneath slightly its structure. In accordance with talk above, here we also shall not discuss the mechanism of this phenomenon. The each individual block of observed structures at every scale of observations has typical structural peculiarities (inherent only to it), and therefore it is able to identify just this block among of similar to many others. At observing of such structures this fact permit us without particular efforts to estimate direction and velocity of a relative motion of such blocks in structure and its absolute value, if the observing object scale is known. Such example is adducted on a Fig. 5.

**Fig. 5.**

In Fig. 6 - Fig. 10 an examples of **CME** images (of processed by **MMDC**) are represented. These figures demonstrate a revealing by means of flow of solar plasma (by generated as a result of the given process) of a large-scale structures (presumably of the **FM**), which (in author's opinion) actually have provoked researched by us the phenomena.

**Fig. 6.**

**Fig. 7.**

**Fig. 8.**

**Fig. 9.**

**Fig. 10.**

### 4. Skeletal Structures of Protuberances and Sun-Spots

So, the filaments of the **FM**, cooperating (taking root into a body) with a surface of the Sun stimulates its activity in this area. As a result of such activity there can be the **CME**, chromospheric flares, protuberances, and also a sun spots (**SS**s) sometimes appear. Here, it will be pertinently to put forward a hypothesis, that the **SS**s can be also by consequence of presence of the **FM** structures inside the Sun, elements of which at its activity can appear on solar surface. The analysis of topology of these structures has shown the identity of their topology to those which already were revealed in the various phenomena connected to presence of such skeletal structures in a wide range of the phenomena, scales and in various environments [1e].

**Fig. 11.**

**Fig. 12.**

**Fig. 13.**

**Fig. 14.**

Fig. 15.

### 5. Observations of a skeletal structures of the FM on the Sun

The structures observations designated by title of the given paragraph include observations of two types. Some of them demonstrate penetration with the outside into an atmosphere and the Sun body of the **CTS**s. As it turned out, some images of such events demonstrate skeletal structures of filaments parts of the **FM** which in reality are latent from the spectator inasmuch it is located behind the Sun disk border. From the analysis of this type of images follows a transparency for an image coming through thickness of the Sun body exist. If that fact that the analyzed images (in this paper) have been obtained in the spectral range of soft x-ray will taken into account, then this phenomenon will by paradoxical. Some discussion of this problem will be given in one of the subsequent paragraphs below. Here one of such images is submitted on a Fig. 16.

**Fig. 16.**

Other type of observations of the **FM** manifestations is based on that fact that sometimes (at the **MMDC** analysis of a temporal dynamics of the investigated images) some blocks of the dark skeletal structures have a radiuses of rotation around of an axis of rotation of the Sun a little smaller than radius of its disk on breadth of their observation. Such structures sometimes are well visible in the near polar areas of star. One of such images is presented on a Fig. 17.

<u>**Fig. 17.**</u>

### 6. The conclusions

So, the analysis (by means of the **MMDC** which the author have developed and described by earlier) of images of the Sun in range of waves lengths of a soft x-ray has resulted [2] in revealing of fractal the **SSS**. The topology of these structures turned out identical to the same that was revealed and described earlier in a wide range of spatial scales, the phenomena and environments [1e]. The **MMDC** analysis of images of the **SS** has shown that they have three-dimensional structure, of the same topology. Due to that, in paper [2] the hypothesis was put forward: *the SSs are a result of manifestation of blocks of the SSS (of the FM) during of its activity period on solar surface.* It would be too courageous assumption, if at this time the fragments **SSS** which have radiuses of rotation around of a star axis a little smaller than the radiuses of a disk of a star on breadth of their locations would not be revealed. The three-dimensional topology of these revealed blocks was easily recognized, due to the analysis of spatial prospect and knowledge of topology of separate structure blocks of the observed **SSS**, identical to those which already were revealed (earlier) in stereoscopic images of laboratory plasma of Z - pinch [1a]. Thus, the **MMDC** analysis of the Sun images in a soft x-ray has led the author to a hypothesis - *the Sun has internal filamentary structure which can be by manifestation of one of probable forms of the FM,* analog of which for the first time was been offered by B.U. Rodionov [4].

If to accept to consideration this model of **FM** then it may be suggested that a formation of the image of its structures inside a body of the Sun occurs due to a process of oscillations of electronic neutrinos at passage of their flux through filaments of the given structure. Just this flux neutrino bears inside itself the formed image of star bowels. Further passage of this flux neutrino across a magnetic field of latent inside of the **FM** filaments (power of this field can reach value up to $3 \cdot 10^{17}$ Gs) located in the nearest space environment of the Sun can lead to emission of quantums of light. The evaluative calculations demonstrates, that if to take account: i) an initial flux solar neutrino; ii) of length of a wave of registered quantums; iii) power of the above-mentioned magnetic field; iv) size of the **FM** structural blocks (corresponding generation) which radius is a little more than radius of solar disk; v) time of an exposition of the image (which is obtained by a telescope of soft x-ray) of order of minutes tens ; vi) the real square of apertur of this telescope, then the calculated probability of considered process is quite suitable (even at condition of averaging (along some volume) of magnetic field of the **FM**) for registration of the image of the **FM** structure inside of the Sun. Unfortunately at present there are no calculations of probability of this process at passage by the flux neutrino of filaments sequences with the big value of a field inside. Most likely, the taking into account of strong heterogeneity of magnetic field along a trajectory of the flux neutrino will lead to substantial growth of probability of considered process.

In the given paper the author had not the purpose to carry out the analysis of above mentioned model of the **FM** and calculations of a probabilities of the possible processes resulting in registration of the image of interiors of our heavenly luminary in the certain range of lengths of waves, and wanted to show only, that the analysis of the off-the-shelf images of the Sun (obtained in the range of soft x-ray) carried out by it, can carry much more information, than it may be expected. Moreover it can spill a light onto physics neutrino.

Really, at carrying out estimations of frequency of radiation as by-product it turns out (at the established parameters of observation) what the restriction which is imposed on rest mass of the electronic neutrino should satisfy to an inequality $m_v \leq 5.1$ eV.

Thus, already now we have a neutrino astronomy which allows us to look into bowels of stars and galaxies, to observe their internal structures and to study processes taking place inside them, to reveal skeletal structures of the **FM** and to study their properties, as inside stars, so in their environment. All this can give a new push into researching of space objects, understanding of processes of their formation, and searching a new energy sources taking place in the universe.

# Figures

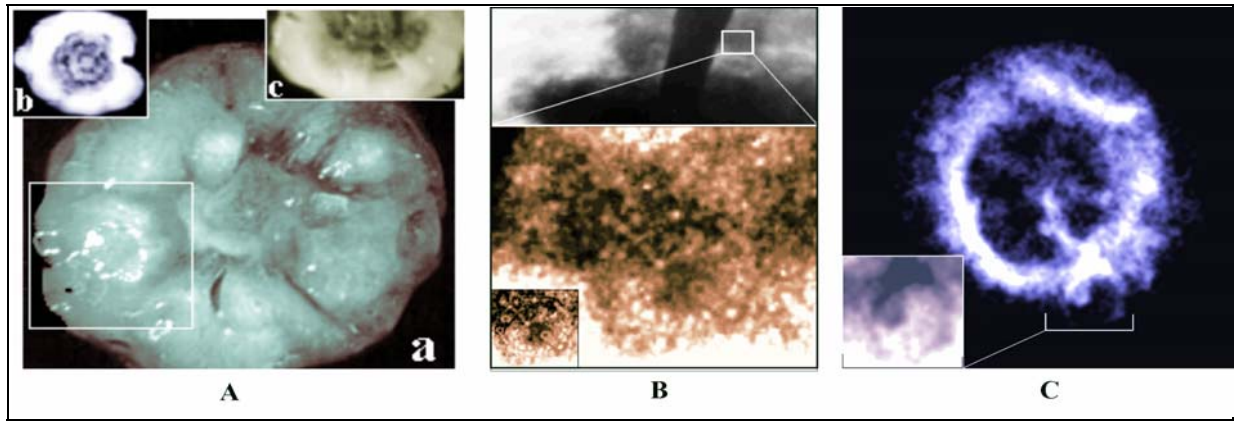

**Fig. 1.** The cartwheel-like structures at different length scales are presented here. **A)** Big icy particles of hail of diameter 4.5 cm (a) [4a], 5 cm (b) [4b], and 5 cm (c) [4a]. The original images are taken from site Ref. [4]. The frame in the left lower part of the image (a) is contrasted separately to show an elliptic image of the edge of the radial directed tubular structure. The entire structure seems to contain a number of similar radial blocks. A distinct coaxial structure of the cartwheel type is seen in the central part of image (b). Image (c) shows strong radial bonds between the central point and the «wheel». **B)** Top section: A fragment of the photographic image [5] of a massive tornado of estimated size of some hundred meters in diameter. Bottom section: A fragment of the top image shows the cartwheel whose slightly elliptic image is seen in the center. The cartwheel seems to be located on a long axle-tree directed down to the right and ended with a bright spot on the axle's edge (see its additionally contrasted image in the left corner insert on the bottom image). **C)** «A flaming cosmic wheel» of the supernova remnant E0102-72, with «puzzling spoke-like structures in its interior», which is stretched across forty light-years in Small Magellan Cloud, 190,000 light-years from Earth (.../snrg/e0102electricbluet.tif [6]). The radial directed spokes are ended with tubular structures seen on the outer edge of the cartwheel. The inverted (and additionally contrasted) image of the edge of such a tubule (marked with the square bracket) is given in the left corner insert (note that the tubule's edge itself seems to possess a tubular block, of smaller diameter, seen on the bottom of the insert). Thus, the cosmic wheel's skeleton tends to repeat the structure of the icy cartwheel up to details of its constituent blocks.

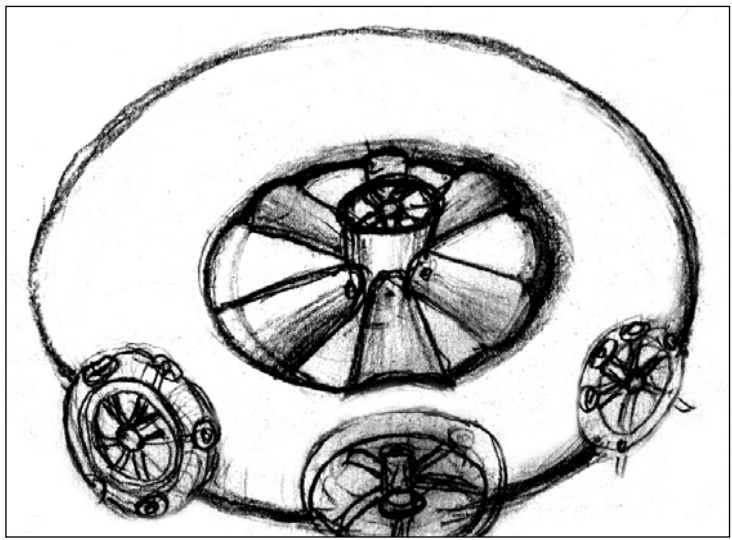

**Fig. 2.** The schematic image of structures such as "cartwheel" is given here. Spokes of a wheel on an outside from its rim show the similar structure, but of the smaller size, i.e., the topology of the given structure shows the tendency to self-similarity.Thus, the cosmic wheel's skeleton tends to repeat the structure of the cartwheel itself up to details of its constituent blocks as in the icy cartwheel.

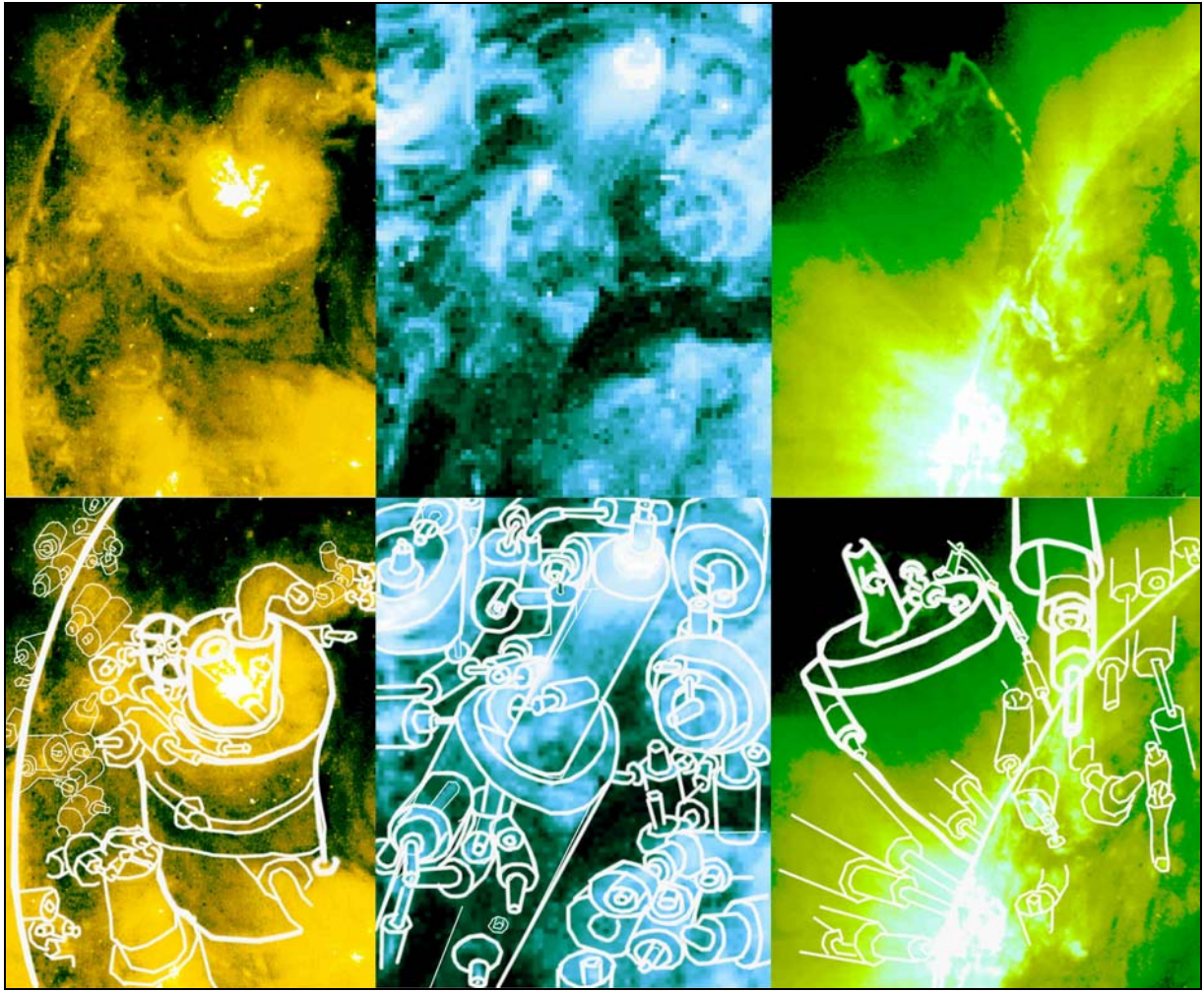

**Fig. 3. i**) In the left, it is given a fragment of the Sun image, obtained in X-ray range, λ = 284 Å (SOHO). The width of image is ~ 3.6 $10^{10}$ cm. The light border of solar disk is visible at the left. The multi-layer, vertically oriented **CTS** of a telescopic type, with maximal diameter ~ 1.8 $10^{10}$ cm is visible in the center. The bright coaxial tube with diameter ~ 2 $10^9$ cm which is located inside of putted forward part of the **CTS** with diameter ~ 3.6 $10^9$ cm is seen very well. The diameters of two neighbor tubes which are fixed in each other have relation ~ 1.6-1.8. Others the **CTS** leaving abroad of a solar disk are visible also. **ii**) In the middle a fragment of the image of the Sun in x-ray area λ = 171 Å (SOHO) is given. Width of figure is ~ 1.25 $10^{10}$ cm. They are visible the **CTS** blocks in various positions and their joints by means of same **CTS** blocks, but the smaller sizes. In the center it is seen inclined to the right from a vertical on a corner in 30° the **CTS** in diameter ~ 5 $10^9$ cm from center of which the telescopic the **CTS** with length ~ 7 $10^9$ cm comes forward. At a butt-end of this structure the diameters of telescopic tubes are visible. Ratio of diameters near by enclosed tubes it is equal ~2. The butt-ends of such tubes reveal the **CTS** of such blocks. **iii**) A fragment of the Sun image in x-ray area λ = 195 Å (SOHO) is represented in the right part of figure. The edge of a solar disk is seen on the right. Diameter towering above a surface of the Sun **CTS** has diameter ~ 2 $10^{10}$ cm and it towers above a surface of our star on height up to ~3-4 $10^{10}$ cm. It is traced its complex the **CTS**.

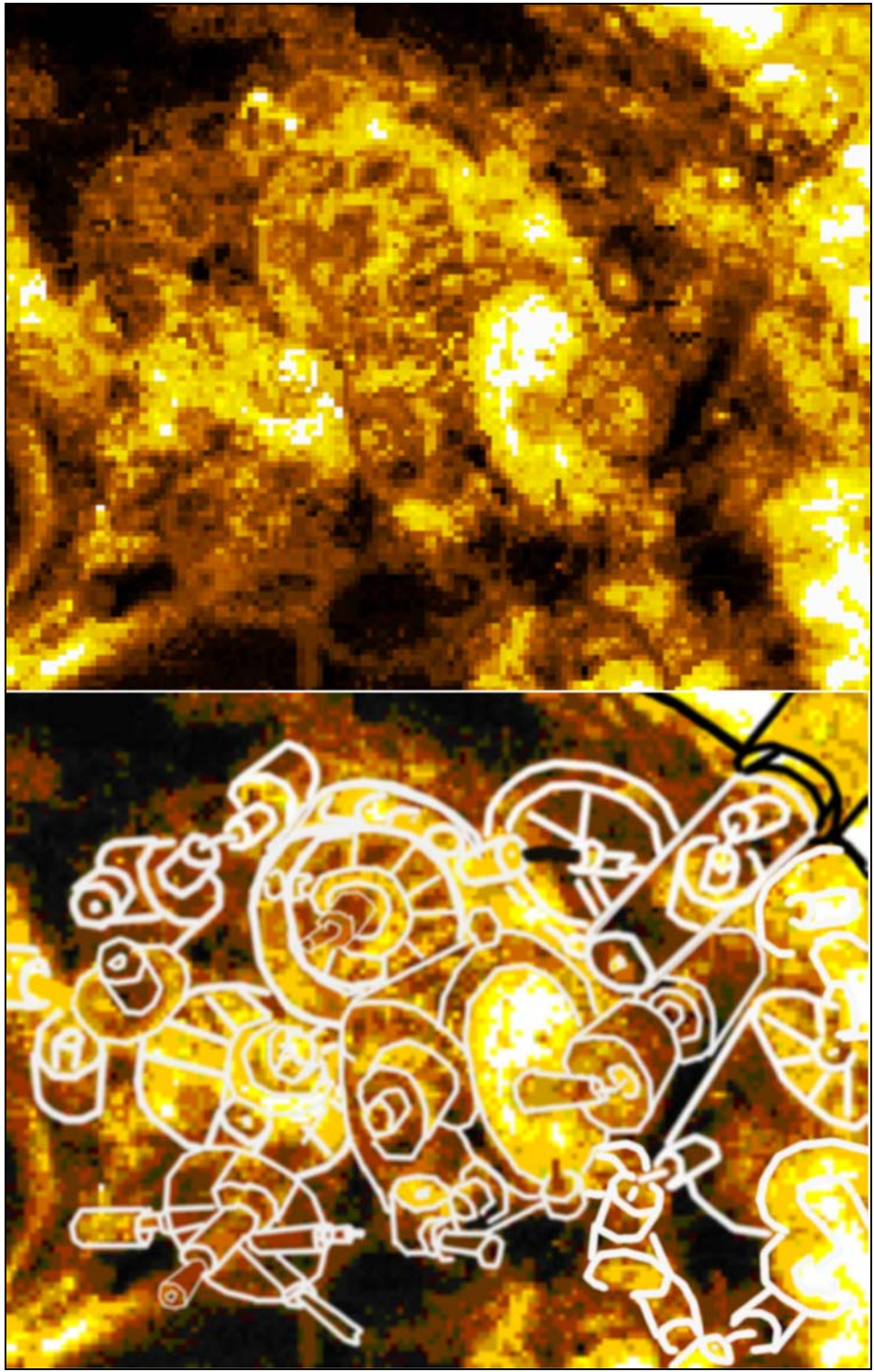

**Fig. 4**. A fragment of the image of the Sun, received in a radio range. Width of the image is ~ $10^{10}$ cm. The border of solar disk is visible on the right, overhead. In the center the disk-shaped the **CTS** with diameter of ~$3 \cdot 10^9$ cm and with an axis oriented to the right and resting on another the CTS (d ~ $10^9$ cm and L ~ $3 \cdot 10^9$ cm), which is located to it orthogonally is visible. Hardly higher of the center and downwards along a figure diagonal it is revealed the **CWS** with diameter ~ $4 \cdot 10^9$ cm and ~ $1.8 \cdot 10^9$ cm and with the well defined a radial spokes which are showing their **CTS** near a rim of the **CWS**. Diameter of the spokes is ~ $1.7 \cdot 10^8$ cm. They are observed also dark **CTS**s of identical topology. The **CTS** (leaving abroad of the solar disk) is visible in a right top corner orthogonally to bordure.

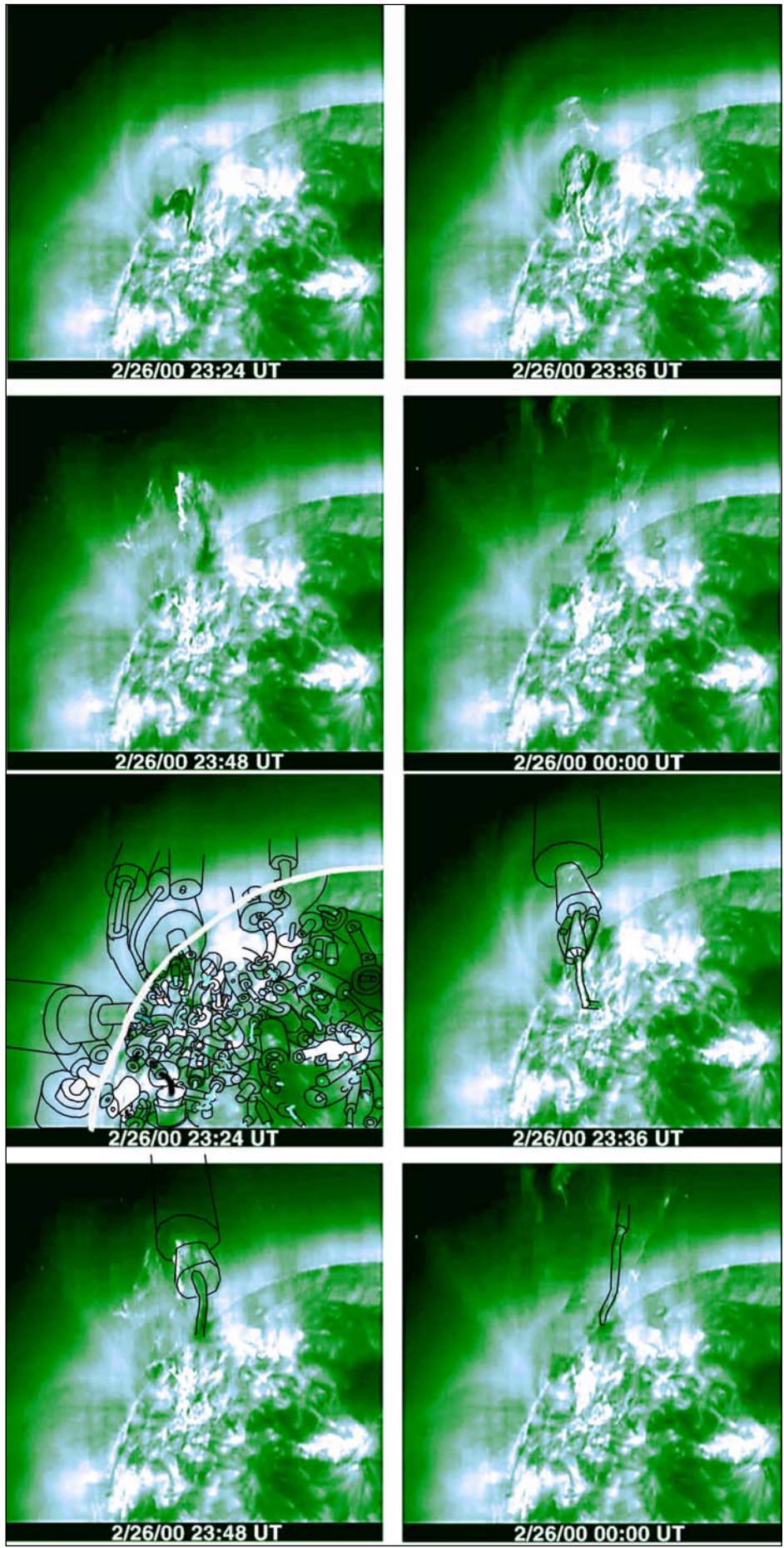

**Fig. 5**. A consecutive fragments of active area of the Sun, λ = 195 Å (SOHO), which are obtained through time intervals in 12 minutes and its schematic images are represented here. Explosion on a surface - result of penetration deep into chromospheres of a large **CTS** (of filament) from the outside. In 2-nd picture the central part of a penetrating filament and further sequence of development of the script of the given event is precisely visible. Here velocities of penetration filament and its movements after explosion are easily estimated, which is ~ $10^7$-$10^8$ cm/s. During all this process any block on the Sun surface has not changed neither it's own an outlines, nor the locations. Therefore, it is possible to tell, that the given perturbation has not led to abyssal processes inside a star, i.e., there was no nuclear explosion inside of the Sun.

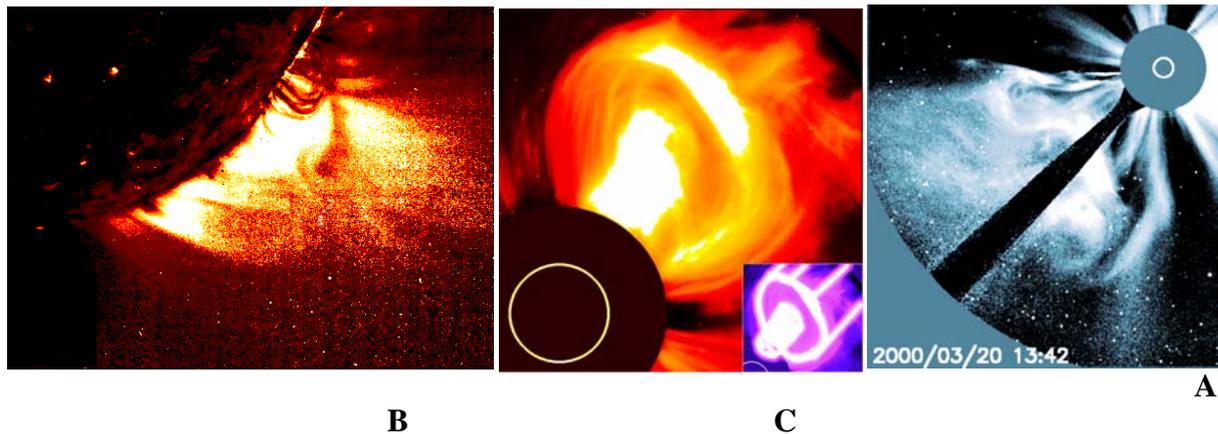

                             **B**                             **C**                       **A**

**Fig. 6**. **A**) A fragment of the image of the Sun $\lambda = 304$ Å (SOHO). Width of a picture is ~ 8 $10^{10}$ cm. The **FM** filament with **CTS** is seen, which is penetrating in body of the Sun (by its a left edge and vertically oriented from below). The stream of plasma caused by this penetration illuminate from beneath of a butt-end of the **FM** filament which stimulated revolting in the Sun atmosphere. The radial connections at a butt-end of this filament are traced. Diameter of this filament ~ $6.5 \cdot 10^{10}$ cm, i.e., it is hardly more than radius of the star. **B**) The x-ray's image of corona mass ejection at explosion on the Sun (LASCO). A schematic representation of the image is given in a window below. On the right along a diagonal the **CTS** with diameter ~ 5 $10^{11}$ cm is visible. The light circle below in the left corner gives scale of the Sun. The schematic image given **CTS** of the **FM** filament is given below on the right in a window. **C**) Emission of weight from the Sun (LASCO). Width of figure is ~ $2.5 \cdot 10^{12}$ cm. On a diagonal downwards from the Sun (a light circle) the conic butt-end of a large filament, cooperating with the Sun is seen. It is well traced the multi-layered **CTS** of it. They are well visible filaments with the **CTS** on a lateral surface of this filament. Here, the light circle – the Sun.

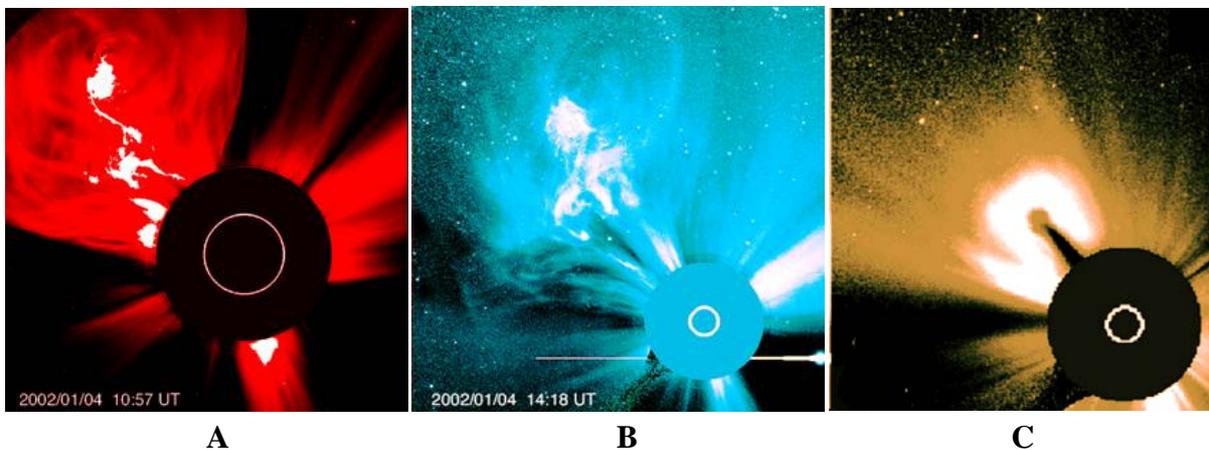

                         **A**                           **B**                        **C**

**Fig. 7**. The images **(A)** and **(B)** are images of the same phenomenon which are showing its development in of time and confirming above suggested hypothesis. The fragment (C) of Fig. 3 gives structure at beginning of this process where a frontal part of the penetrating FM filament is seen. The fragments (A) and (B) of this picture reveal the beginning of response of a star onto this penetration (emission of plasma and highlighting of the **CTS** of this **FM** from below. The diameter of this **CTS** is equal to about 10 diameters of the Sun, and the butt-end of it is seen almost on a diagonal from above down and to the right. **C**) A picture of the **CME** of the Sun (LASCO) is presented here. Width of figure is ~ $6.6 \cdot 10^{13}$ cm. A conic butt-end of a large filament, cooperating with the Sun is well seen along a figure diagonal upwards from the Sun (a light circle). The structure of this filament has a coaxial-tubular design. It is well seen a conical continuation of lateral surface of this filament.

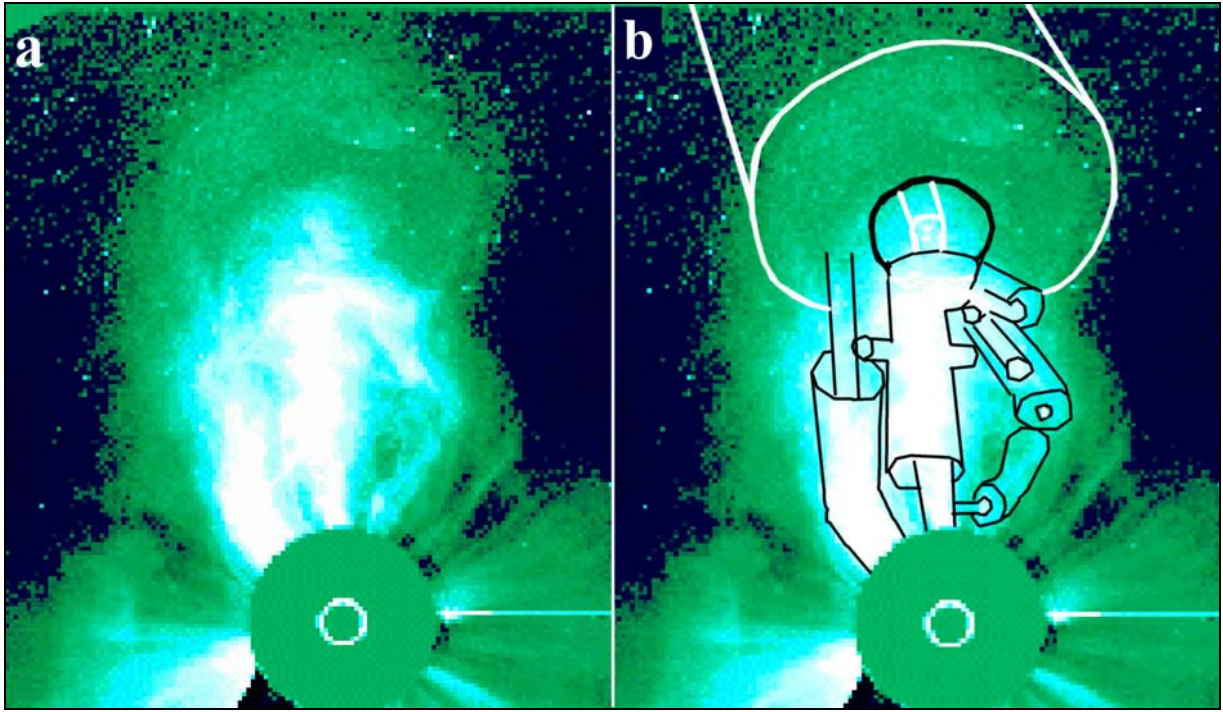

**Fig. 8.** The width of figure ~ **2.5 $10^{11}$ cm** (LASCO) is well looked through a flow and visualization of filament of **FM** which is directed upwards from the Sun and afar from the spectator. Diameter of the filament ~ **$10^{12}$ cm**. Diameters of the central parts of it's **CTS**, directly cooperating with the Sun ~ of diameter of the star and more. Communication of this filament with similar, which is directed across is traced.

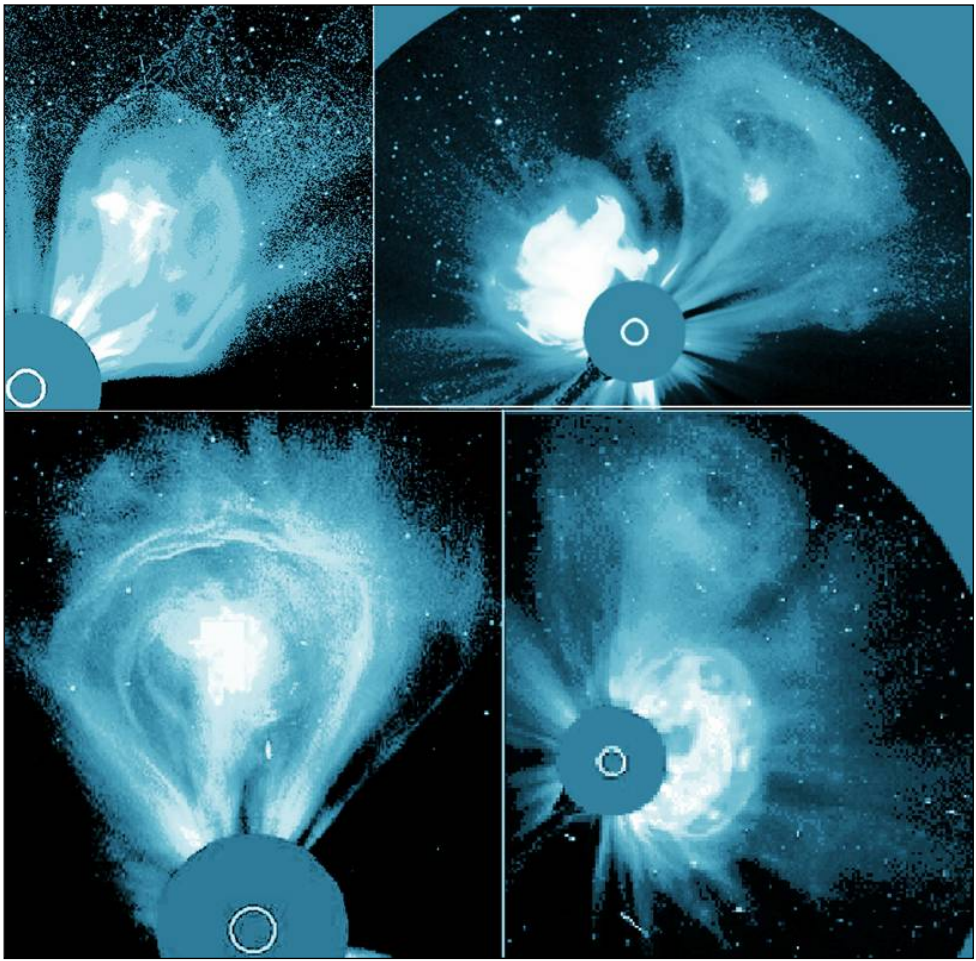

**Fig. 9.** CT filament are revealed from database LASCO-C2. The scale is defined by the size of the Sun which is marked by a white circle. Scale of these large-scale filament ~ **$10^{12}$ cm.** Filament are multilayered and have radial communications as a **CTS** spokes. Development of **CME** occurs during of some hours that corresponds to plasma velocity up to ~ **$10^8$ cm /c.**

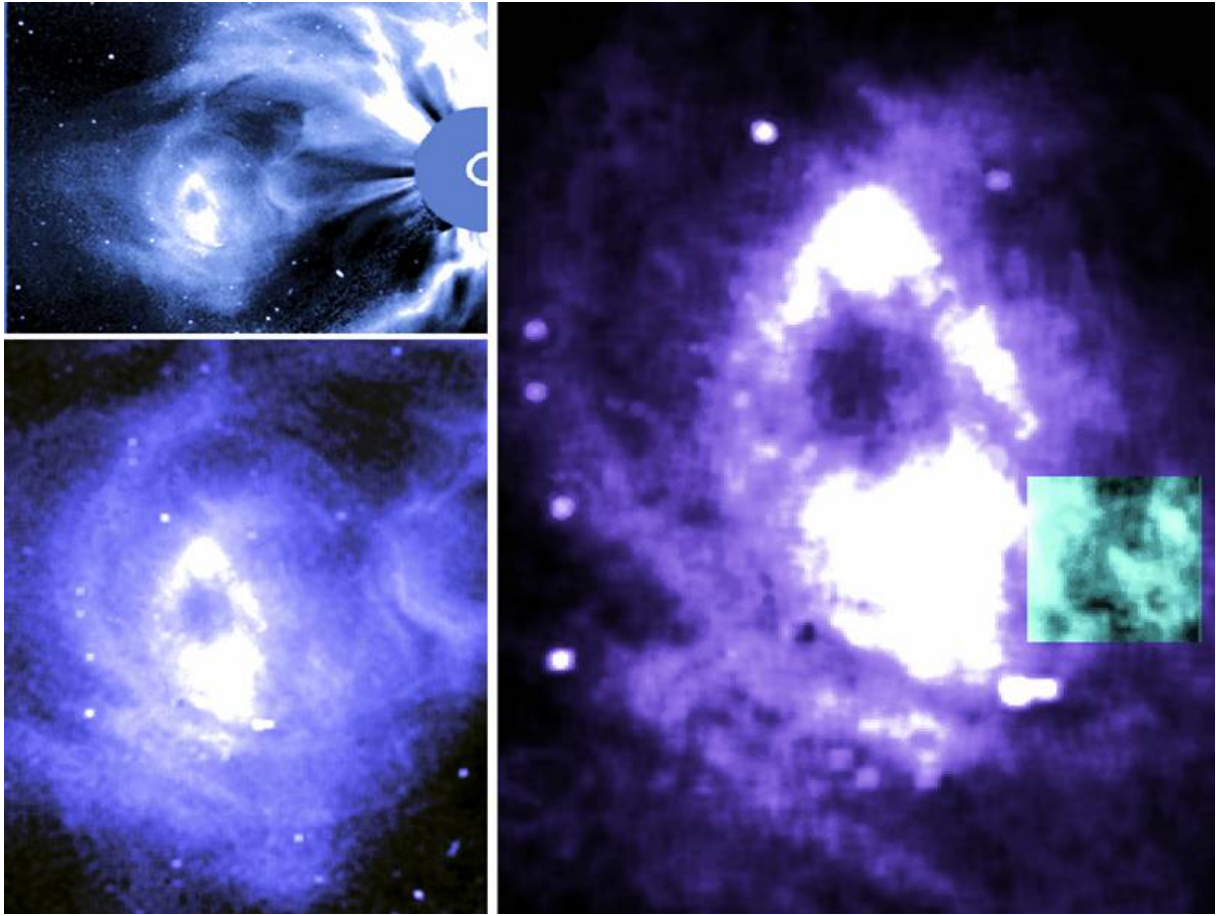

**Fig. 10.** Revealed at the **CME** of the Sun (LASCO) a design of the **FM** filament, which has provoked explosion is represented here. The image of solar coronal mass ejection with the solar disk overshadowed and indicated with a white circle is given on upper window at the left. The magnified image of the front of the jet reveals a wheel-like structure with straight radial bonds between thick central axis and rim of this cartwheel is presented in window lower. The magnified and additionally contrasted image of the wheel-like structure; the window (seen on the right) differs by contrast and color to show the continuity of the wheel radial spoke. The elliptic images with the small central point (which are seen on the front edge of this spoke and in the central point of the wheel, and in the right-hand radial spoke) suggest all these formations to posses a tubular (and even coaxial) structure. So, the radial connections in the form of filaments with the **CTS** which passing through the cartwheel rim of construction of the **CTS** are precisely revealed and show the self-similar structure.

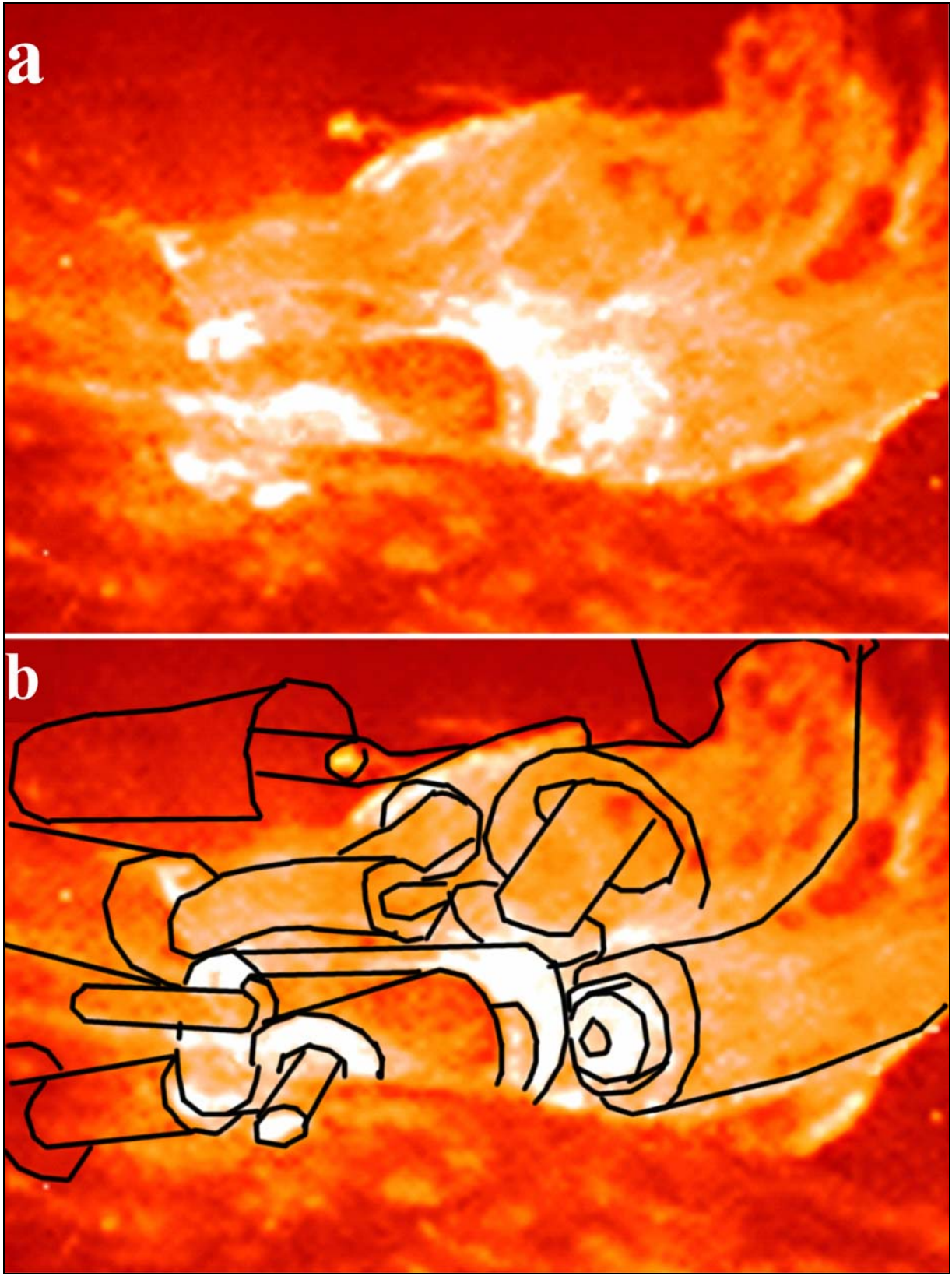

**Fig. 11**. The filament of **FM**, cooperating (taking root into a body) with a surface of the Sun causes its activity in this area. At the bottom of the **CTS**s of this protuberance is shown. Width of figure is ~ 3.8 $10^9$ cm. Diameters of the **CTS**s are ~ 7 $10^8$ cm.

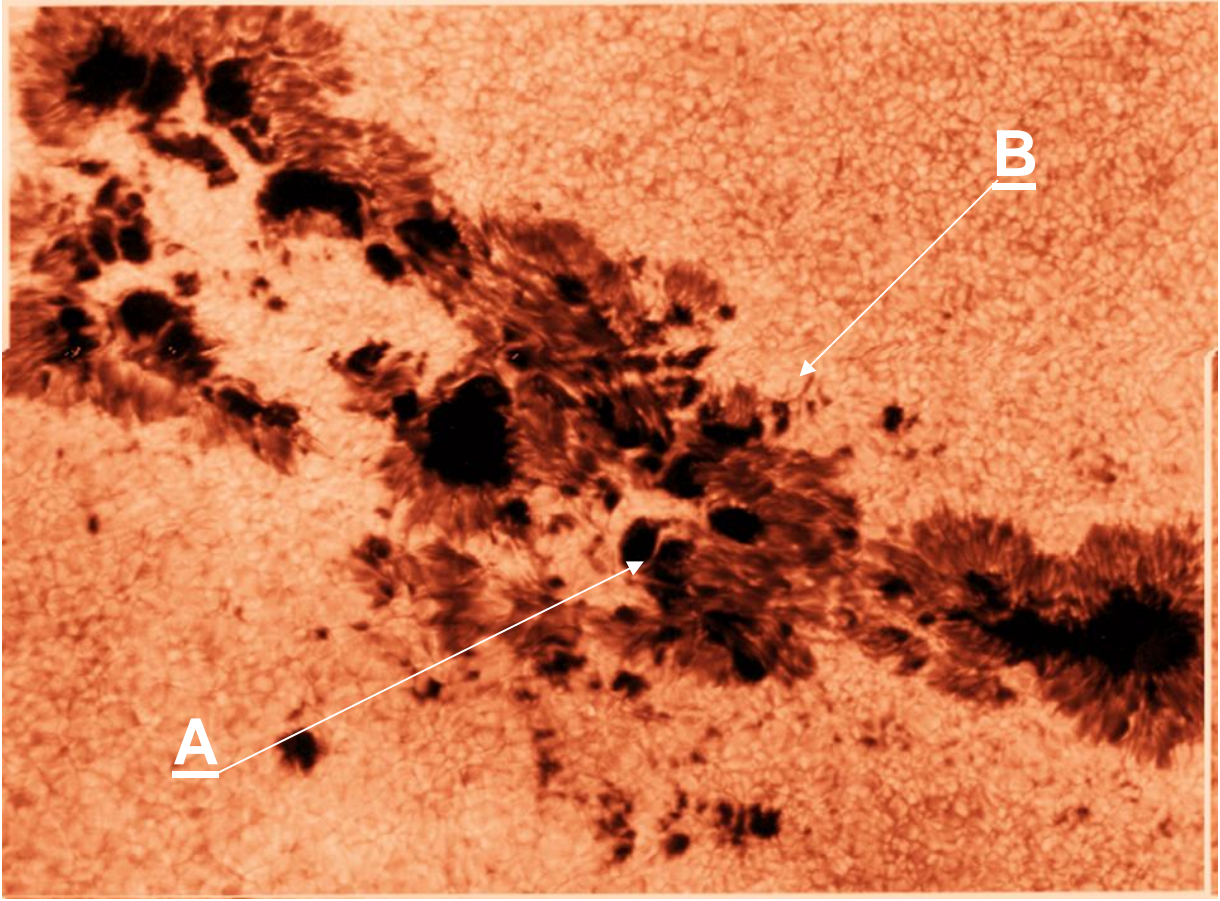

**Fig. 12.** The active area of the Sun surface with a chain of **SS**s, the structures almost of all of them have a three-dimensional topology. Arrows designate areas which are in more detail submitted on the following figures.

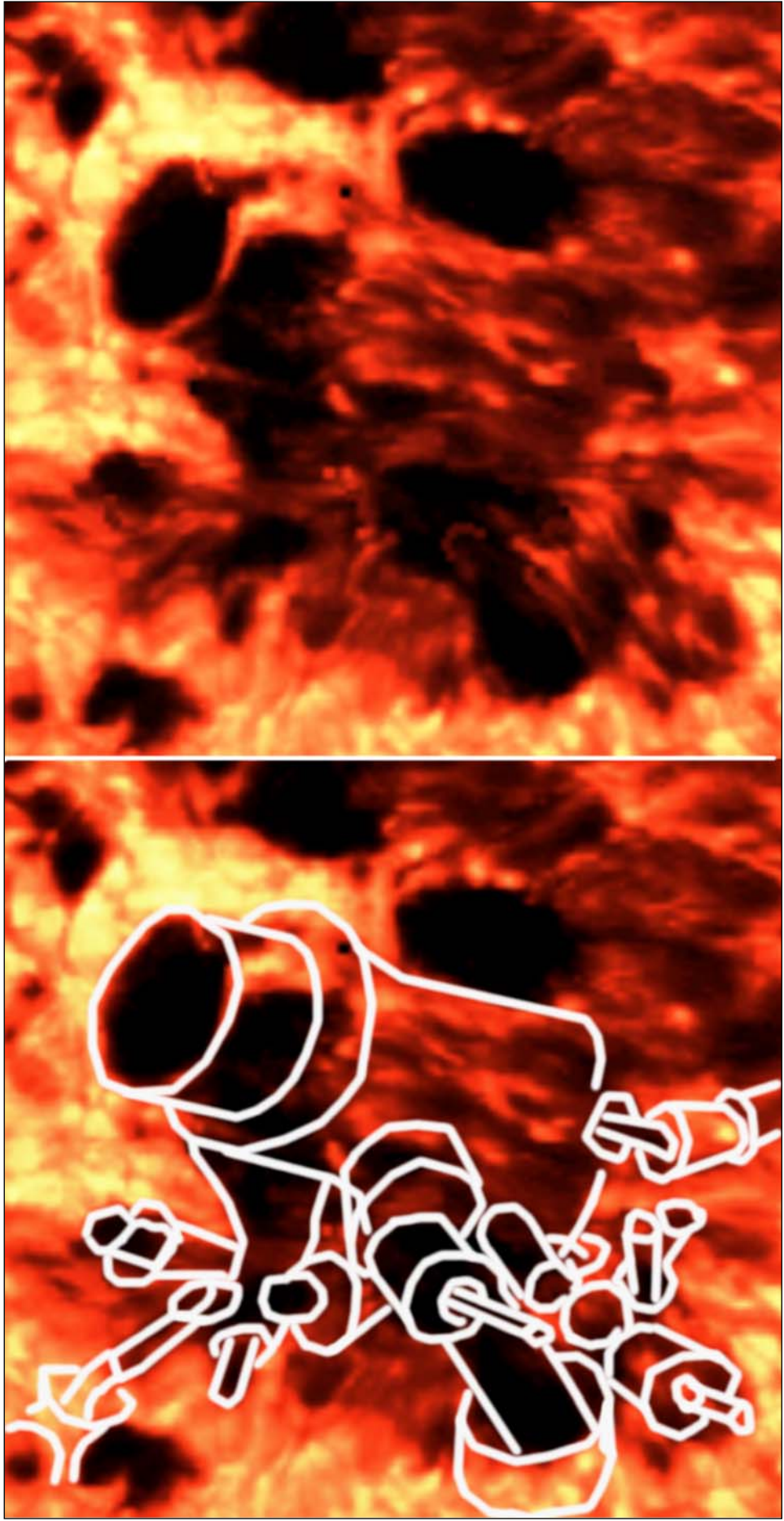

**Fig. 13.** The fragment of active zone (along arrow "<u>A</u>" on Fig.9) of the Sun surface and its schematic representation are given on this figure. It is visible, that the **SS** represent **CTS** of the **FM** push out from bowels of our star on its surface. The topology of this structure is the same, as well as in observable before in dust deposit of tokamak T-10. Tridimentionality of these structures, their inter-lacings and connection is precisely traced. Diameter structure is $\sim 2 \cdot 10^9$ cm. The structure is situated along the left diagonal of figure and butt-end of a cylindrical tube by length $\sim 8 \cdot 10^9$ cm, located along this diagonal.

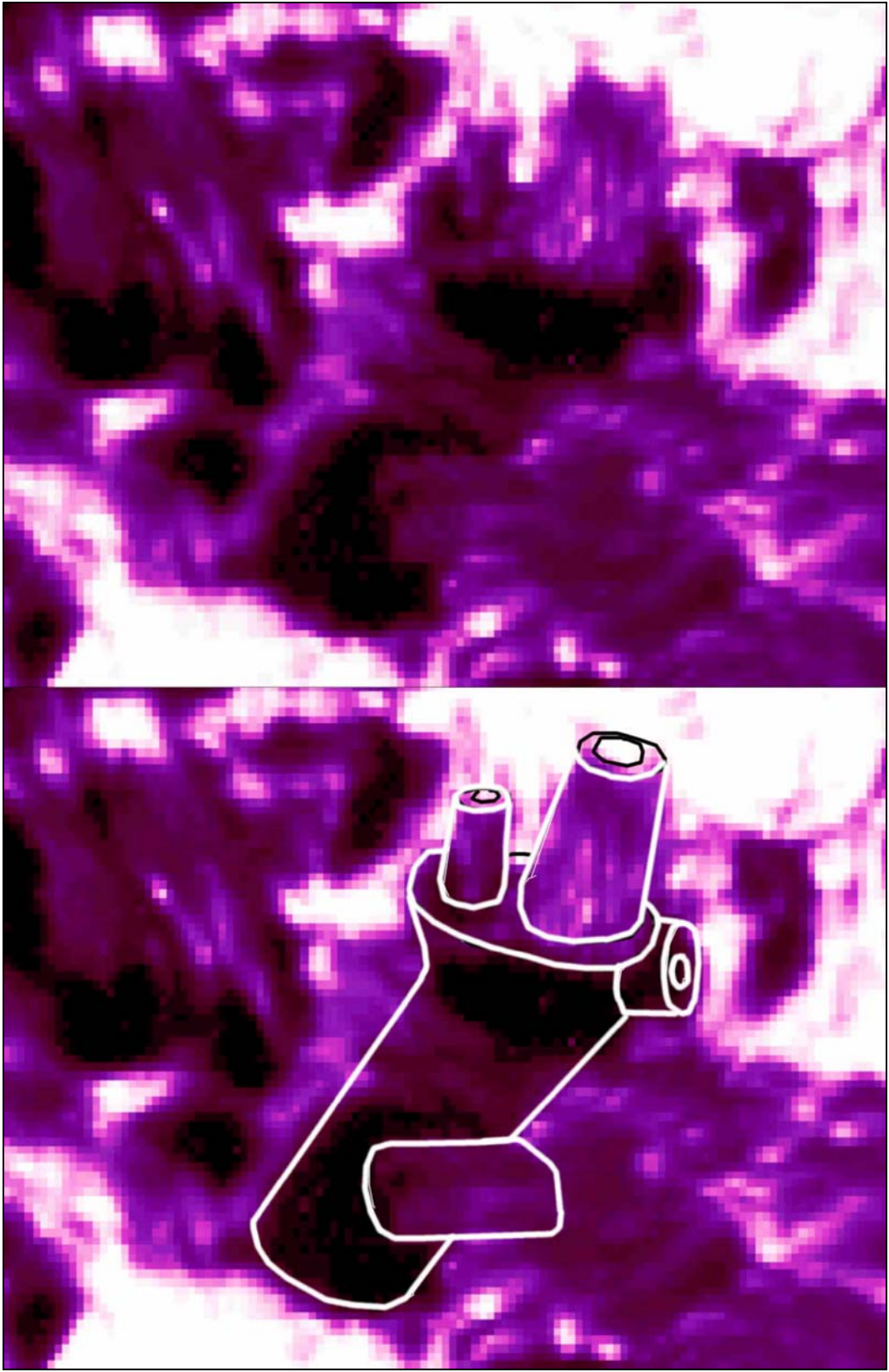

**Fig. 14.** The another fragment of active zone of the Sun surface (along arrow "<u>B</u>" on Fig.9) and its schematic representation are given here. The image of this **SS** which represents a dark volumetric **CTS** with diameter ~ 2 $10^9$ cm and which is towered above a surface of the Sun onto height up to ~ $10^{10}$ cm is given. Connections of this structure with similar others of ones are seen.

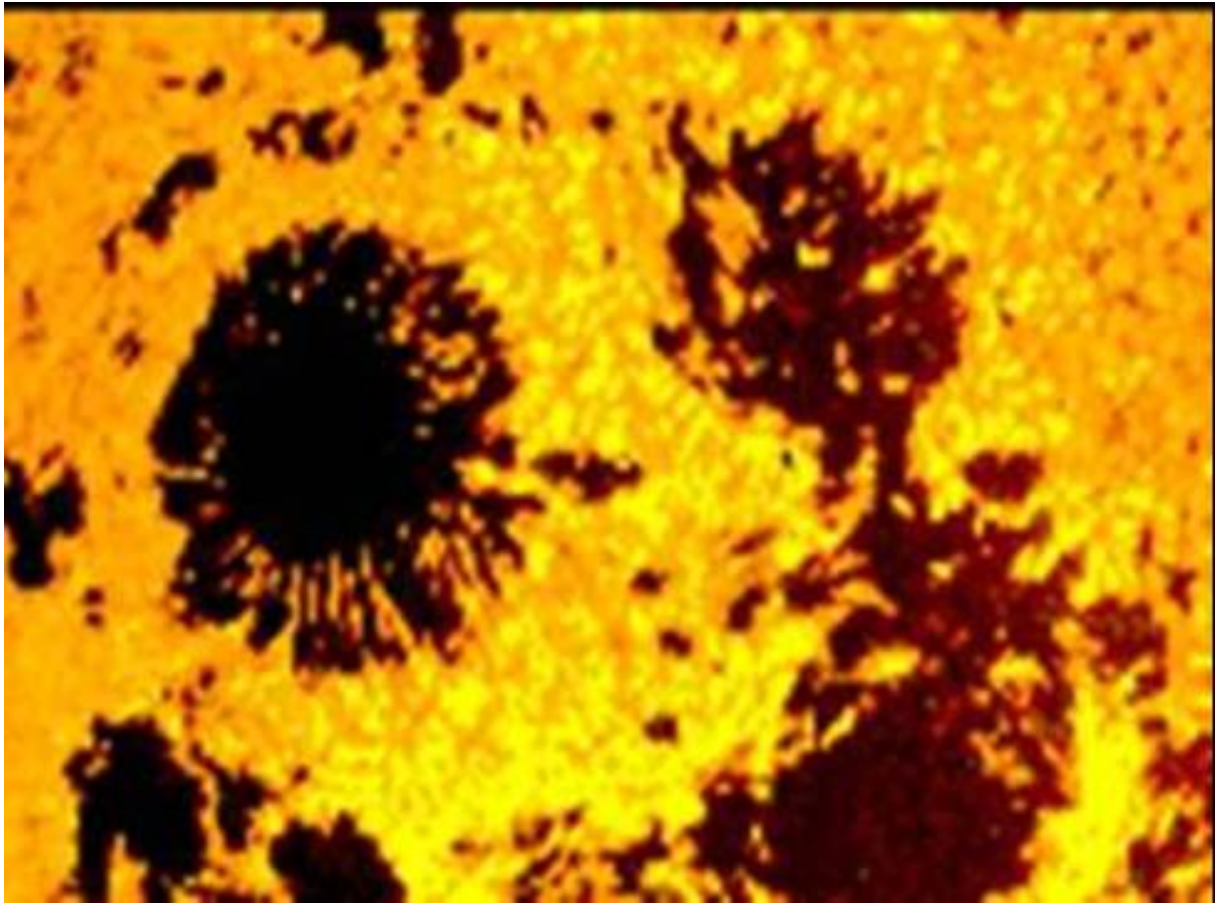

**Fig. 15.** Here the image of an active zone of the **SS** which form **CWS** is submitted. The center of this **CWS** is a big **SS** which is located almost in the centre of this figure. This **CWS** above sink a little into a star body. Communications between separate **SS** are here too traced. It is seen that these **SS** everything are also a three-dimensional structures which is assembled of tubular identical blocks.

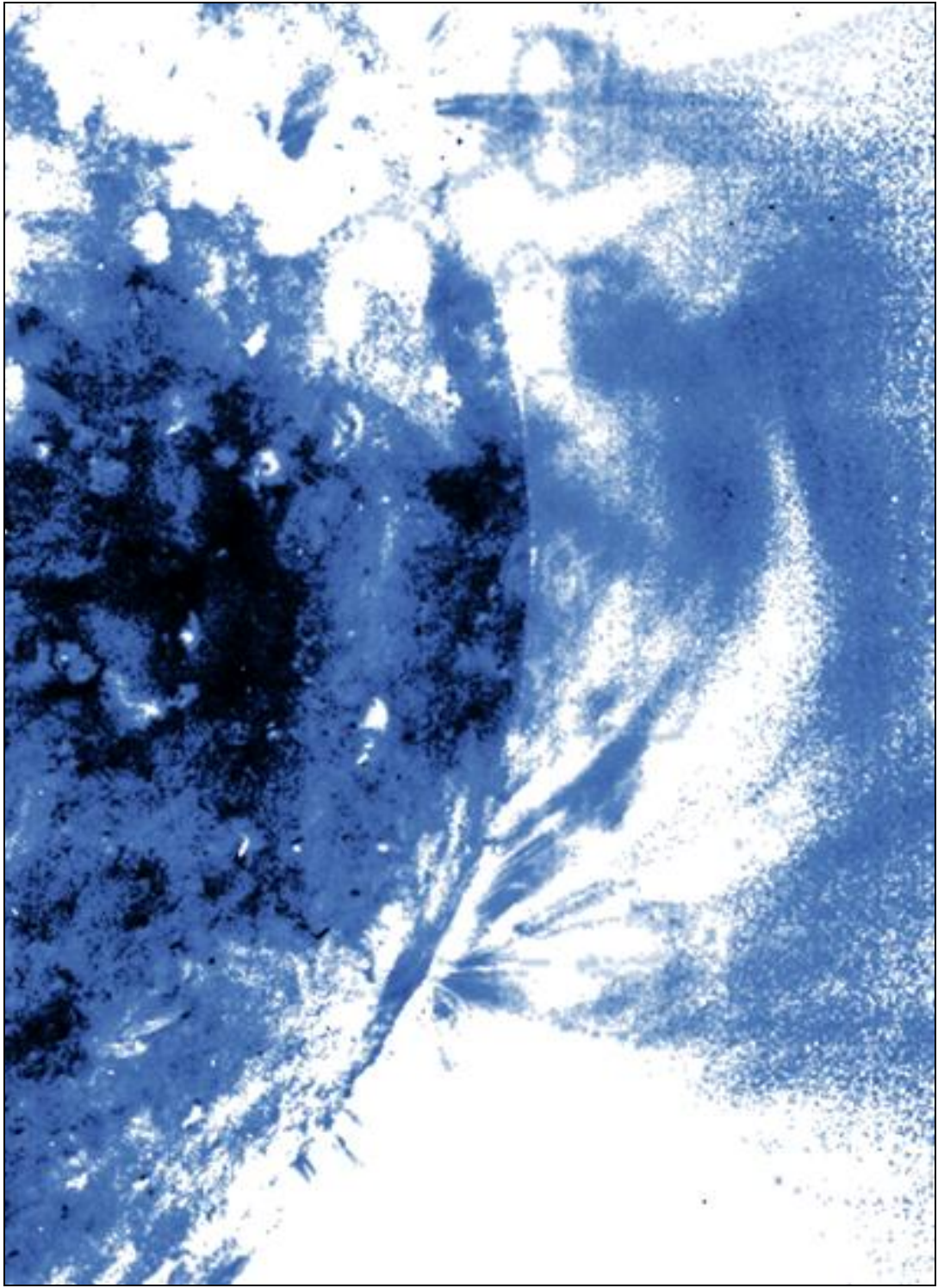

**Fig. 16**. A fragment of the image of the Sun ($\lambda = 304$ Å, SOHO), processed by means of **MMDC**. Height of figure is $\sim 7.5\ 10^{10}$ cm. The filament by diameter at its butt-end near of border of the Sun disk $\sim 4.8\ 10^{10}$ cm is given here. This filament runs into a body of the star across with some inclination on the spectator and a little downwards. It is traced prolongation of this filament structure into the left from border of our star. It is seen, the large-scale filament is made of filaments the smaller sizes. They are precisely traced their **CTS**s. Butt-end of this filament and structure of its part latent by a body of the star on this figure as though are seen through thickness of substance of a body of the Sun. This can seem strange if to remember, the observation is carried out in lines of soft x-ray.

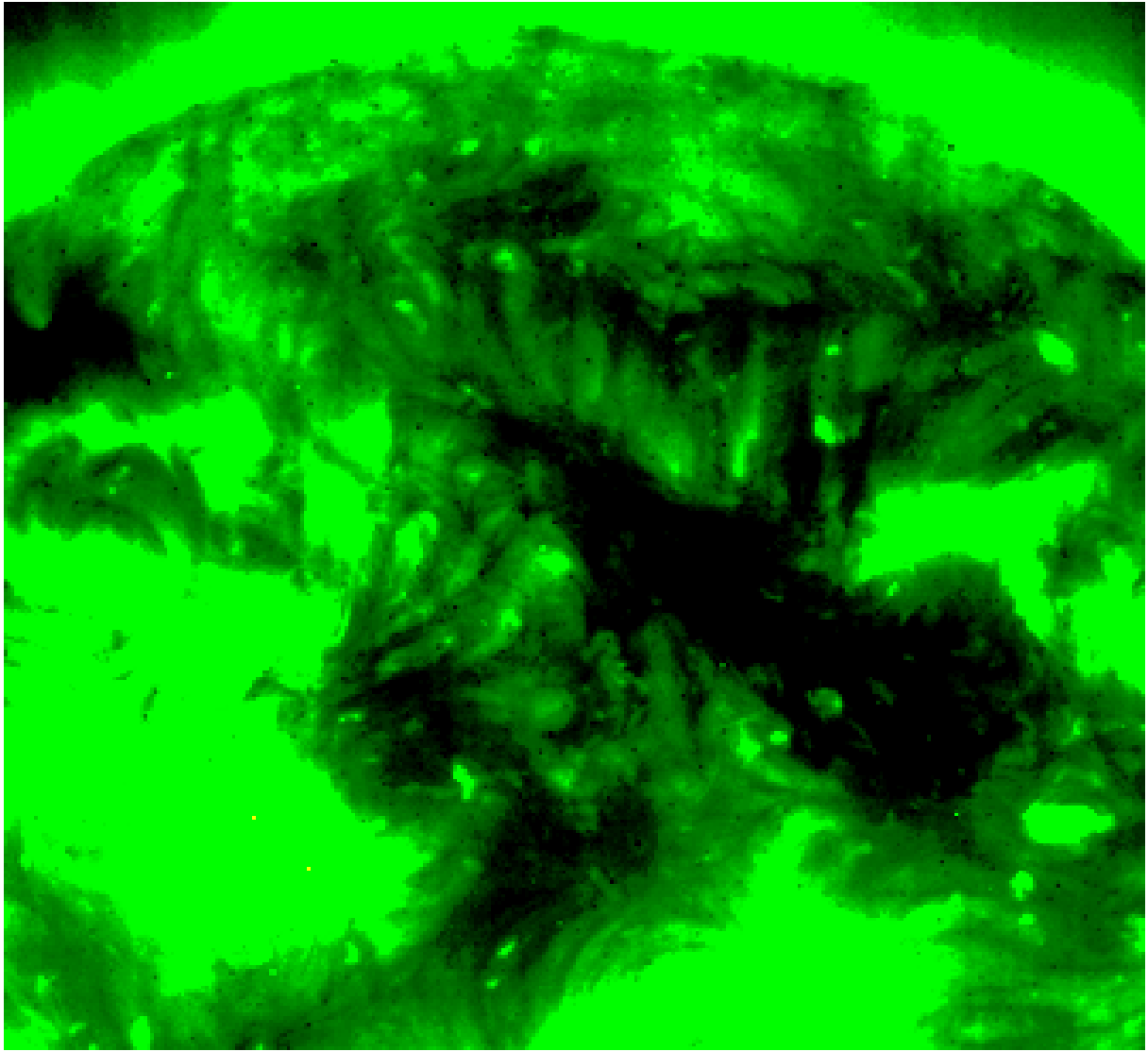

**Fig. 17**. This is a fragment of the Sun image obtained in the soft X-ray at the highest phase of its activity (after the **MMDC** processing of this fragment was resulted). The edge of a solar disk is seen overhead. Here it is revealed the structure (which, presumably, is complexly weaved of the **FM** filament and collected of self-similar blocks the **CTS**s, but of the smaller size). Diameter of the **CTS** is ~ $3.3 \ 10^{10}$ cm, diameter of blocks of which it is collected is ~ $10^9$ cm, diameter of the block going out from the center of its butt-end is ~ $8 \ 10^9$ cm. This structure has radius of rotation around of an axis of rotation of the Sun of smaller radius of a solar disk on breadth of its observation.